\documentclass[useAMS,usenatbib,referee]{mn2e}
\usepackage{epsfig}
\usepackage{graphicx}
\usepackage{lscape}

\usepackage{color}

\title[Population study for $\gamma$-ray
millisecond pulsars]
{Population study for $\gamma$-ray
pulsars: II Millisecond pulsars}
\author[J. Takata, Y. Wang and K.S. Cheng]{J. Takata \thanks{E-mail:
takata@hku.hk}, Y. Wang \thanks{E-mail:yuwang@hku.hk}  and  K.S. Cheng\thanks{E-mail:hrspksc@hkucc.hku.hk}\\
Department of Physics, University of Hong Kong,
Pokfulam Road, Hong Kong
}
\begin{document}

\date{}

\pagerange{\pageref{firstpage}--\pageref{lastpage}} \pubyear{2010}

\maketitle

\label{firstpage}

\begin{abstract}
The population of $\gamma$-ray emitting millisecond pulsars (MSPs) is studied by using
Monte-Carlo techniques. We simulate the Galactic distributions of the MSPs,
and   apply the outer gap model for the  $\gamma$-ray emission
 from each simulated MSP.
We take into account  the dependence of the observed $\gamma$-ray flux
 on the viewing angle and inclination angle, which is the angle
 between the rotation axis and the magnetic axis, respectively. Using
 the sensitivity of the six-month long observation of the $Fermi$
telescope and radio sensitivities of existing pulsar surveys, 9-13 radio-selected and 22-35
 $\gamma$-ray-selected pulsars are  detected within our simulation. 
 The statistical properties of the simulated population are consistent 
with the $Fermi$ observations. Scaling the observed sensitivity 
$\propto \sqrt{T}$, where $T$ is the
 length of observation time, the present model predicts
that over the 5-year mission $Fermi$  would detect 15-22
radio-selected $\gamma$-ray MSPs,  and 95-152 $\gamma$-ray-selected  MSPs. Our
simulation also predicts that about 100 (or 200-300) $\gamma$-ray MSPs with a flux larger $F\ge 10^{-11}~\mathrm{erg/cm^2 s}$
(or $5\times 10^{-12}~\mathrm{erg/cm^2 s}$) irradiate the Earth.
 With the present sensitivities 
of the radio surveys, most of them are categorized as
$\gamma$-ray-selected pulsars, indicating that most of the $\gamma$-ray
MSPs have been missed by the present $Fermi$ observations.
We argue that the Galactic $Fermi$ unidentified sources located
at  high latitudes should
be dominated by MSPs, whereas the sources in the galactic plane
are dominated by
radio-quiet canonical pulsars.  We want to emphasize that the predicted number of
radio-loud $\gamma$-ray MSPs depends on the sensitivities of radio surveys and
 that is  can be increased, for
example, from 15-22 to 26-37 if the radio sensitivity is improved by a factor of 2.
\end{abstract}

\section{Introduction}
\label{intro}
The millisecond pulsars (MSPs), which have  a rotation period 
$P\sim 0.002-0.1$~s and
 a stellar magnetic field  $B_s\sim 10^{8-9}$~G, are classified
as a different generation from the canonical pulsars,
which have $P\sim 0.03-10$~s and $B_s\sim 10^{12-13}$~G.
The $Fermi$ LAT has detected $\gamma$-ray
emissions from about  60~pulsars in just  two year observation, including
 9 millisecond pulsars (Abdo et al. 2010a,b,c, 2009a,b;
Saz~Parkinson et al. 2010).  Furthermore,
the detection of  radio MSPs
 from about 20 unidentified $Fermi$ point
sources (e.g. Ray 2010; Caraveo 2010; Ranson et al. 2011) has been reported.
The MSPs are now recognized as  one of the major populations of the
 Galactic $\gamma$-ray sources.

The particle acceleration and high-energy $\gamma$-ray radiation
processes in the pulsar
magnetosphere have been studied  with  the polar cap  model
(Ruderman \& Sutherland  1975; Daugherty \& Harding 1982, 1996),
the slot gap model (Arons 1983; Muslimov \& Harding  2003; Harding
et al. 2008) and the outer gap model (Cheng, Ho \& Ruderman 1986a,b;
Hirotani 2008; Takata, Wang \& Cheng 2010a) respectively.
All models have  assumed that the charged particles are accelerated by
the electric field along the magnetic field lines, and that the
accelerating electric field  arises in the charge
deficit region, where the local
charge density  deviates   from the
Goldreich-Julian charge density (Goldreich \& Julian 1969).
The polar cap model  assumes that the acceleration region is near the stellar
surface.  On the other hand,  the outer gap model and the slot gap model
assume a strong acceleration region extending to the  outer magnetosphere.
For the canonical $\gamma$-ray pulsars,  several access of observational
evidence have been  proposed for  the outer magnetospheric origin
of the $\gamma$-ray emissions  (Aliu et al. 2008; Abdo et al., 2009c).
Because the strength of the  magnetic field at
 the outer magnetosphere of the MSPs is
similar to that of canonical pulsars, $B\sim 10^{5-6}$~G,  it is
expected that if the $\gamma$-ray emissions of the MSPs  originates from the
outer magnetosphere,
 the emission characteristics (e.g. pulse profile and
spectrum) will be  similar to those of  the canonical pulsars.

Romani \& Watters (2010) and Watters \& Romani (2011) studied  
the pulse profiles of the canonical  pulsars
observed by $Fermi$. They computed the pulse profiles predicted
 by the outer gap model and slot gap model,  and  argued statistically 
that the outer gap geometry is more
  consistent with the observations than the slot gap geometry.
  Venter, Harding \& Guillemot (2009) fit the pulse profiles of the
 $Fermi$ detected MSPs with the geometries predicted by
the different high-energy emission models.  They
found that the pulse profiles  of  two out of
eight   millisecond pulsars cannot be fitted by the geometries of either 
 the outer gap or  the slot gap  models.  They proposed
 a pair-starved polar cap model, in which the multiplicity of the pairs
is not high enough to completely screen the electric field above the
polar cap,  and the particles are continuously accelerated up to high
altitude over the entire open field line region. Thus we see that
 the $\gamma$-ray emission mechanism and the emission site 
in the MSP's magnetosphere have not been satisfactorily explained up to now.

The increase of population of detected  $\gamma$-ray emitting pulsars allows us
 to perform a more detailed population study of the high-energy emissions
from the pulsars.  A comparison between the simulated and observed
distributions of the various pulsar
 characteristics (e.g. rotation period and  $\gamma$-ray flux) will
be useful to test the high-energy emission model.
 With  the canonical $\gamma$-ray
pulsars,  Takata, Wang and Cheng (2011) have studied the population
predicted by the
outer gap model. They predicted that with the sensitivity of the six-month long 
$Fermi$ observation, about 100 $\gamma$-ray emitting canonical
pulsars can be detected,  suggesting that the present observations have missed
many $\gamma$-ray emitting pulsars.   For the MSPs,
 Story, Gonthier and Harding (2007) have studied the
population of $\gamma$-ray MSPs with
the slot gap accelerator models and predicted the $Fermi$ observations.
 They predicted $Fermi$ will detect 12 radio-loud and 33-40 radio-quiet $\gamma$-ray MSPs.

The population study of the MSPs is also  important for an understanding  the
unidentified $\gamma$-ray sources detected by $Fermi$. In the $Fermi$
first source catalog, there are  several hundred  unidentified steady point
sources (Abdo et al. 2010b). Takata et al. (2011) have argued
that the  $Fermi$ unidentified sources located in the high Galactic latitudes 
  cannot be explained by the galactic distribution of the canonical pulsars.
The millisecond pulsars are possible candidates for the high
Galactic $Fermi$ unidentified sources.

In this paper,  we develop a Monte-Carlo study of the population
of the $\gamma$-ray emitting MSPs predicted by the outer gap model. In particular, we only study the population of MSPs in the Galactic field (not in 
 the globular clusters). 
Following the previous studies (e.g. Story et al. 2007; Takata et al. 2011),
we will perform a Monte-Carlo simulation for the Galactic population
of the MSPs (section~\ref{simulation}).   In section~\ref{gemission},
we will discuss our $\gamma$-ray emission model, including
dependence of the $\gamma$-ray flux on the viewing geometry.
In section~\ref{result}, we present the results of our simulation.
In particular, we will compare the simulated  population with the $Fermi$ 
six-month long observation (section~\ref{comparison}).
 In section~\ref{gamsp}, we discuss the possibility of
 MSPs as the  origin of the Galactic $Fermi$ unidentified sources. We will
discuss the results of our simulation in section~\ref{conclusion}.

\section{Monte-Carlo simulation}
\label{simulation}
\subsection{Galactic population}
\label{motion}
In this study, we assume that all MSPs are born through the so called recycled
process in the low-mass binary systems,  with a birth rate of $10^{-6}\sim 10^{-5}~\mathrm{yr^{-1}}$ (Lorimer et al. 2005;
Lorimer 2008). The birth location
 is determined  by the  spatial distributions proposed by
Paczynski (1990),
\[
\rho_R(R)=\frac{a_{R}\mathrm{e}^{-R/R_{\mathrm{exp}}}R}{R^2_{\mathrm{exp}}},
\]
\begin{equation}
\rho_z(z)=\frac{1}{z_{\mathrm{exp}}}\mathrm{e}^{-|z|/z_{\mathrm{exp}}},
\end{equation}
where $R$ is the axial distance from the axis through the Galactic
centre  perpendicular to the Galactic disk and  $z$ is the distance from
the Galactic disk. In addition, $R_{\mathrm{exp}}=4.5$~kpc,
$a_R=[1-\mathrm{e}^{-R_{max}/R_{exp}}(1+R_{max}/R_{exp})]^{-1}$ with $R_{max}=20$
~kpc and $z_{exp}=200$~pc (Story et al. 2007).

To obtain the current position of each simulated MSP, we
solve the  equations of motion from its birth to the current time.
The equations of motion are given by
\begin{equation}
 \frac{dR^2}{dt^2}=\frac{v_{\phi}^2}{R}-\frac{\partial
 \Phi_{tot}}{\partial R},
\label{eqr}
\end{equation}

\begin{equation}
 \frac{dz^2}{dt^2}=-\frac{\partial \Phi_{tot}}{\partial z},
\label{eqz}
\end{equation}
and
\begin{equation}
Rv_{\phi}=\mathrm{constant}.
\end{equation}
Here $v_{\phi}$ is the azimuthal component of the velocity,  $\Phi_{tot}=\Phi_{sph}+\Phi_{dis}+\Phi_h$ is the total gravitational potential, where 
$\Phi_{sph}$, $\Phi_{dis}$ and $\Phi_{h}$ are the spheroidal, the disk
and the halo components of the Galactic gravitational potential, and  are given
by
\begin{equation}
\Phi_i(R,z)=-\frac{GM_i}{\sqrt{R^2+[a_i+(z^2+b_i^2)^{1/2}]^2}},
\end{equation}
where $i=sph$ and $dis$. For the spheroidal component, $a_{sph}=0$, $b_{sph}=0.277$~kpc and
$M_{sph}=1.12\times 10^{10}M_{\odot}$.
For the disk component, $a_{dis}=3.7$~kpc, $b_{dis}=0.20$~kpc,
and $M_{dis}=8.07\times 10^{10}M_{\odot}$,  while for the halo
component 
\begin{equation}
\Phi_{h}(r)=-\frac{GM_{c}}{r_c}\left[\frac{1}{2}\ln
 \left(1+\frac{r^2}{r_c^2}\right)+\frac{r_c}{r}\tan^{-1}
\left(\frac{r}{r_c}\right)\right],
\end{equation}
where $r_c=6.0$~kpc $M_c=5.0\times 10^{10}M_{\odot}$, respectively  (c.f. Burton \& Gordon 1978; Binney \& Tremaine 1987; Paczynski 1990). The Lagrangian in units of energy
per unit mass is given by
\begin{equation}
L=\frac{v^2(R,z,\phi)}{2}-\Phi_{tot}(R,z),
\end{equation}
where $v$ is the velocity of the pulsar.

For the distribution of the initial velocities of the MSPs,
we assume  a Maxwellian
distribution with a characteristic width of $\sigma_v=70$~km/s (c.f.
Hobbs et al. 2005), namely,
\begin{equation}
\rho_v(v)=\sqrt{\frac{\pi}{2}}\frac{v^2}{\sigma_v^3}
\mathrm{e}^{-v^2/2\sigma_v^2}.
\end{equation}
For the azimuthal component, the  circular
 velocity due to the Galactic gravitational
potential field at the birth position of the MSPs is taken into account, and 
it is  calculated from
\begin{equation}
v_{circ}=\left[R\left(\frac{\partial\Phi_{sph}}{\partial R}+
\frac{\partial\Phi_{dis}}{\partial R}+\frac{\partial\Phi_{h}}{\partial R}
\right)\right]^{1/2}.
\end{equation}

\subsection{Pulsar characteristics}
It is widely accepted that the MSPs are so called recycled pulsars, which were
 spun up by accretion of the matter from the low mass companion star.
 The rotation period of the newly born MSP is related to the history of the
accretion onto the neutron star. In particular,
its rotation period in the accretion stage may be related to the
equilibrium spin period,
$P_{eq}\sim 1.7B_8^{6/7}\dot{M}_{15}^{-3/7}R_6^{18/7}M_{1.4}^{-5/7}$, where
$B_8$ is the neutron star magnetic field in units of $10^8$~Gauss,
$\dot{M}_{15}$ is the accretion rate in units of $10^{15}$~g/s, $R_6$ is
the neutron star radius in units of $10^6$~cm and $M_{1.4}$ is the neutron
 star mass in units of 1.4 solar mass (Frank, King \& Raine, 2002).
However, the description of the
transition from an accretion powered to the rotation powered phase is not well understood
due to the complexities in the description of the interaction
between the magnetosphere of a neutron star and its accretion disk.
Furthermore,  the transition, for which a rapidly decreasing
accretion rate is required (e.g. Jeffrey 1986),
 may be facilitated by different processes between
the long and the short orbital period systems.  For the long orbital period
systems, a red giant companion can detach from its Roche lobe as its
envelope is exhausted. On the other hand, for the short orbital period 
systems,  a possible mechanism leading to the sudden decreas 
in the accretion flow is
the operation so called ``propeller effect'' (Campana et al. 1998),
or dissolution of the disk by  $\gamma$-ray irradiation in the
 quiescent stage (Takata, Cheng \& Taam, 2010b). The initial
period of the MSPs will depend on the characteristic of the orbital motion and/or
 the termination mechanism of the accretion flow.

In the present study, we assume that
all MSPs are not directly produced  by supernova explosions,
but that they are born by the recycled process in a low-mass binary system.
 On this assumption, the true age of the MSPs,
defined by the time since the neutron star was born in the supernova
explosion, is different from the spin down age (i.e. $P/2\dot{P}$), and  from
the age defined by activating  the rotation powered activity.  Because the time
scale of
the recycled process, including the decay of the neutron star's 
magnetic field and the accretion process, is not
understood well, it is  very difficult to
 calculate the present period and the magnetic field strength of MSPs from their initial distributions and
the true age of MSPs. 

\begin{figure}
\begin{center}
\includegraphics{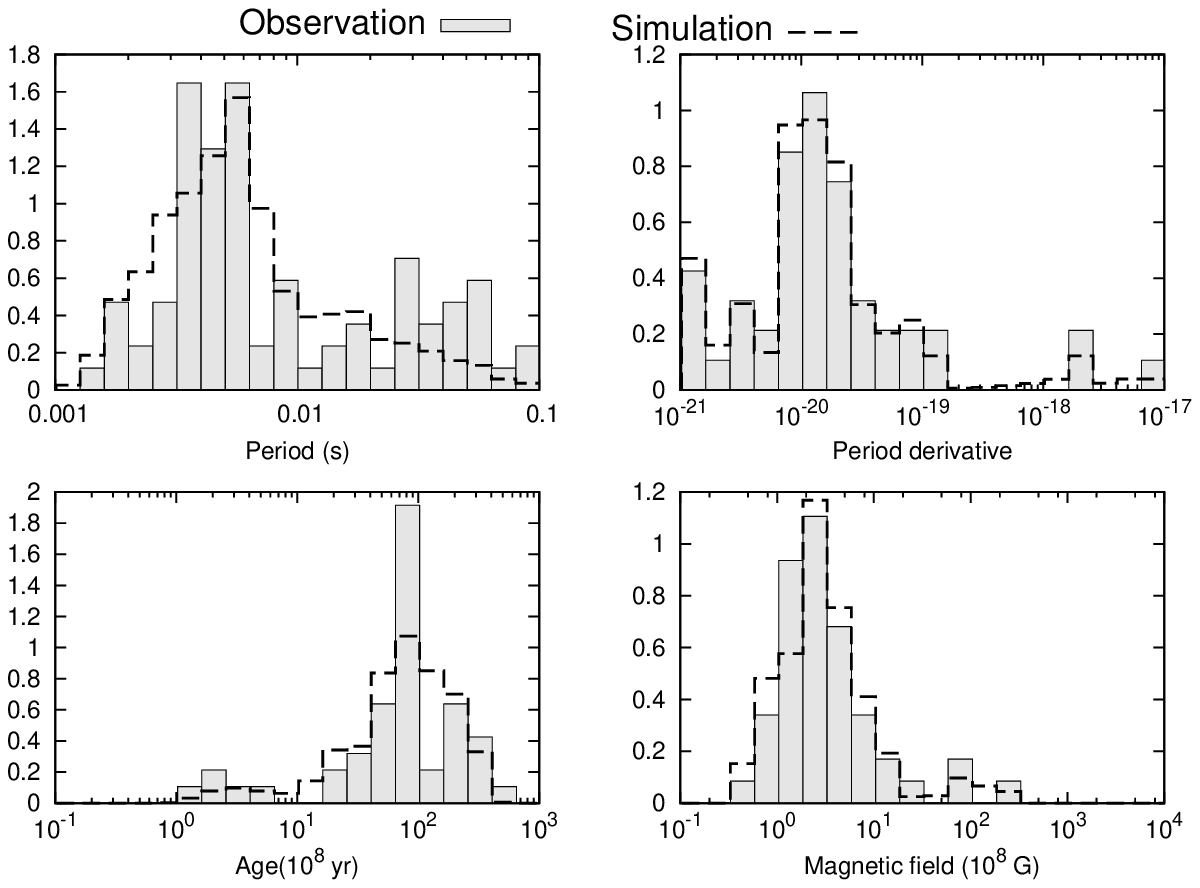}
\caption{The observed distributions (shaded histograms)
of the various characteristics for radio MSPs. The dashed lines
show the distributions applied for the present simulation.
The observed distributions are for the intrinsic values (i.e. without
Shklovskii effect) of the pulsar parameters.  }
\label{robs}
\end{center}
\end{figure}

Allowing for  uncertainties, we use
the observed distribution of the radio MSPs to assign  ``current''
 pulsar characteristics (e.g. rotation period, magnetic field) for each
simulated MSP, instead of modeling the initial distribution.
In the usual  Monte-Carlo studies for the Galactic population of 
pulsars (e.g. Story et al. 2007; Takata et al. 2011), the initial period
is assigned for each simulated pulsar and then the current position and
rotation period are obtained.
In the present study, on the other hand, we (1) generate the simulated
MSPs  with a constant birth rate over $10$~Gyr,  (2) obtain the current position
as described in section~\ref{motion} and (3) assign the pulsar parameters
following the observed  distributions.
Specifically,  we assign  the period time derivative ($\dot{P}$)
and the stellar magnetic field ($B_s$) for each simulated MSP
following the observed
$\dot{P}-B_s$ distribution, where we use the intrinsic value after removing the
  Shklovskii effect (see below). In this paper, we denote as  $B_s$ the strength
of the magnetic field  at the magnetic equator, namely,
$B_s=3.2\times 10^{19}\sqrt{P \dot{P}}$. 
The current rotation period and the spin down age
of simulated MSPs are calculated from
\begin{equation}
P_{-3}=0.97B_8^2\dot{P}_{-20}^{-1}~\mathrm{ms}
\end{equation}
and
\begin{equation}
\tau=1.5\times 10^9 P_{-3}\dot{P}_{20}^{-1}~\mathrm{yr}
\label{age}
\end{equation}
respectively (Lyne \& Graham-Smith, 2006).
Here $P_{-3}$ and $\dot{P}_{-20}$ are the rotation period in units
of 0.001~s and the period time derivative in units of $10^{-20}$, respectively.
The distributions, which are used in the present simulation, are represented for
the various characteristics of the simulated MSPs  in
 Figure~\ref{robs} by the dashed  lines.
Figure~\ref{robs} also presents the observed distributions
(shaded-histograms, Manchester et al. 2005).

Shklovskii (1970) argued that the Doppler shift resulting from the
transverse motion of a pulsar makes a positive contribution to
the pulsar's period time derivative as (Manchester 1999),
\begin{equation}
\dot{P}_s=\frac{P\mu^2d}{c},
\label{shk}
\end{equation}
where $\mu$ is the proper motion of the  MSP. For the MSPs 
whose  period time derivative
is extremely small, the Shklovskii effect may significantly increase
the period time derivative. For canonical
pulsars, on the other hand, the period time derivative is large, e.g.  $\dot{P}\sim 10^{-13}-10^{-15}$, which is much greater than  that caused by
the Shklovskii effect.

We note that the present procedure can be applied if
the Galactic distribution of the MSPs   does not depend on the
 age of the rotation powered MSPs (and spin down age); that is,
the Galactic MSP's populations  of 1~Gyr and of 10~Gyr MSPs, for example,
 are described by the same distribution.  With the present assumption
that all MSPs are born through
the recycled process, which has a characteristic time scale completely independent of the spin down age, the 
 Galactic distribution is independent of the distribution of the spin down age.
With a typical velocity of the
observed MSPs ,  $v\sim 70$~km/s,  it is expected that the typical
displacement of MSPs  with an age, $\tau\ge 100$~Myr,
becomes larger than the size of the Galaxy. However, with  the slow velocity,
the MSPs remain bound to the Galaxy and hence
their Galactic distribution does  not  depend on the age of
the rotation powered MSPs.
On the other hand, for canonical pulsars (in particular for
the $\gamma$-ray pulsars),
 the Galactic distribution depends on the true age, which is almost
equal  to the  spin-down age. This allow us  to
 calculate the present period distribution
from the initial distribution.

\subsection{Radio emissions}
Using the empirical relations
among the radio luminosity, rotation period, and period time derivative,
 the distribution of the
 radio luminosity  at 400~MHz is expressed by (Narayan \& Ostriker 1990)
\begin{equation}
\rho_{L_{400}}=0.5\lambda^2\mathrm{e}^{\lambda},
\end{equation}
where  $\lambda=3.6[\mathrm{log_{10}}(L_{400}/<L_{400}>)+1.8]$ with
$L_{400}=\eta 10^{6.64}\dot{P}^{1/3}/P^3$, and
$L_{400}$ is the luminosity in units of  $\mathrm{mJy~kpc^2}$.
Here $\eta$ is a scaling factor to adjust the observed distribution, 
and $\eta=1$ for the canonical pulsars.
In the present simulation for the MSPs,
 we find that  $\eta\sim0.05$ can explain the distribution of the observed
radio luminosity of the MSPs.  The radio flux on 
Earth is given by $S_{400}=L_{400}/d^2$, where $d$ is the distance to the MSP.
We scale the simulated 400~MHz luminosity to the observational frequency using
a typical photon index $\sim$1.8 (Kramer  et al. 1998).

We also take into account the beaming effect of the radio emissions.
The half-angle, which is  measured from the magnetic axis,
 of the radio emission cone of the MSPs does not depend on the frequency, and
 is approximately  described by  (Kramer \&  Xilouris, 2000),
\begin{equation}
\omega\sim 5.7^{\circ}P^{-1/2}.
\end{equation}
  This emission can be detected
by  observers with a viewing angle between max($0^{\circ}$,$\alpha-\omega$)
 and min($\alpha+\omega$, $90^{\circ}$),
where $\alpha$ is the inclination angle
between the rotation axis and the magnetic axis.

We use  the ten radio surveys (Molongo 2, Green Band 2 and 3, Arecibo 2 and 3,
Parks 1, 2 and MB, Jordell Bank 2 and Swinburne IL), whose
 system characteristics are listed in Table~1 of Takata et al. (2011)
and the references therein.  To calculate the dispersion measure, we use
 the Galactic distribution of electrons obtained by  Cordes \& Lazio (2002).

\subsection{$\gamma$-ray emission model}
\label{gemission}
\subsubsection{Observed $\gamma$-ray flux}
\label{oflux}
In this paper we apply  the
outer gap accelerator model 
(Cheng, Ho \& Ruderman 1986a,b;  Zhang \& Cheng 2003;
Takata et al. 2010a) for the millisecond pulsars.  In the outer gap,
the electrons and/or  positrons are accelerated up to
a Lorentz factor of $\Gamma\ge 10^7$ by the electric field along
the magnetic field line.
These accelerated particles can emit $\gamma$-ray photons of several GeV 
through the curvature radiation process. Assuming the force balance between the
electric force and the radiation drag force, the Lorentz factor is given by
\begin{equation}
\Gamma=\left(\frac{3R_c^2}{2e}E_{||}\right)^{1/4},
\end{equation}
where $R_c$ is the curvature radius of the magnetic field lines and $E_{||}$ is
 the accelerating electric field. The accelerating electric
field in the outer gap is approximately given
by (Cheng et al. 1986a,b; Cheng, Ruderman \& Zhang 2000)
\begin{equation}
E_{||}(r,z)\sim \frac{B(r)f(r)^2R_{lc}}{R_c} \frac{z}{z_h}
\left(1-\frac{z}{z_h}\right),
\end{equation}
where $R_{lc}=cP/2\pi$ is the light cylinder radius and $f(r)$ is the
 gap thickness divided by the light cylinder radius ($R_{lc}=cP/2\pi$)
 in the poloidal plane. In addition, $z$ is the
height measured from the last-open field line in the poloidal plane, and
$z=z_h$ is the upper boundary of the gap. We note $B(r)f(r)^2\sim$constant 
along the field line  for
the dipole field geometry.

The spectrum  of the curvature radiation emitted by an individual
particle may be written as
\begin{equation}
P_c(E_{\gamma},r)=\frac{\sqrt{3}e^2\Gamma}{hR_c}F(x),
\end{equation}
where $x=E_{\gamma}/E_c$ with $E_c=3hc\Gamma^3/4\pi R_c$ and
\[
F(x)=x\int_x^{\infty}K_{5/3}(t)dt,
\]
where $K_{5/3}$ is the modified Bessel function of  order 5/3.
If the $\gamma$-ray beam  points toward an observer, the observer
will measure a phase-averaged spectrum of
 (Hirotani 2008),
\begin{equation}
\frac{dF_{\gamma}}{dE_{\gamma}}\sim \frac{1}{d^2}\int
N(r_{\xi}) R_c(r_{\xi}) P_c(E_{\gamma},r_{\xi})d A,
\label{flux}
\end{equation}
where $N=B/Pce$ is the particle number density, $r_{\xi}$ represents the
radius from which the emission can be measured by the observer,
 $dA$ is
the cross section of the gap perpendicular to the magnetic field lines.
We assume that the outer gap extends in the azimuthal direction with $\delta\phi\sim
 \pi$~radian.
The integrated energy flux
between 100~MeV and 300GeV can be calculated from
\begin{equation}
F_{\gamma,>100~MeV}=\int_{100MeV}^{300GeV}\frac{dF_{\gamma}}{dE_{\gamma}}dE_{\gamma}.
\label{gaflux}
\end{equation}

 One can show that the $\gamma$-ray flux described by equation~(\ref{gaflux})
 approximately satisfies  $F_{\gamma,>100~MeV} \propto f(R_{lc})^3L_{sd}$,
 where $L_{sd}=4(2\pi)^4B_s^2R^6/6c^3P^4$ is the spin down power.  If
the typical energy of a curvature photon $E_c$ satisfies   $100~\mathrm{MeV}
\le E_c\le 100$~GeV, the power radiated by a single charged particle is
$\int P_cdE_{\gamma}\sim 2e^2c \Gamma^4/3R_c^2$.  Furthermore,
if we estimate the total number of charged particles in the gap
 in terms of values near the light cylinder as $N_{gap}\sim \frac{B(R_{lc})}{Pce}R_{lc}\delta A(R_{lc})$, we obtain
\begin{equation}
F_{\gamma,>100~MeV}\sim \frac{1}{d^2}
\frac{2e^2c \Gamma^4}{3R_{lc}^2}N_{gap}
\sim \frac{f^3 B^2(R_{lc})R^3_{lc}\delta\phi}{d^2P}\propto
\frac{f^3B_s^2R_s^6}{d^2P^4c^3}\propto f^3L_{sd}/d^2,
\end{equation}
where we have used $\Gamma\sim (3R_{lc}^2E_{||}/2e)^{1/4}$,
$E_{||}\sim B(R_{lc})f^2(R_{lc})$ and $\delta A\sim f \delta\phi R_{lc}^2$.
The $\gamma$-ray flux is proportional to the cube of the fractional
gap thickness and the spin down power. 

We also assume as a zeroth order approximation that the gap current is 
of the order of the Goldreich-Julian value over the full width.
 The detailed calculation of the
 outer gap model (e.g. Takata \& Chang 2007) gives the
current distributions in the direction perpendicular to the magnetic field lines.
On the other hand, by fitting the observed $gamma$-ray spectra of all mature pulsars detected by $Fermi$,Wang, Takata \& Cheng
(2010) find
that the total current in the gap is of the order of the Goldreich-Julian
value. We expect  that as long as
the total power of the $\gamma$-ray radiation
from the outer gap accelerator is concerned, the uniform current
distribution with the Goldreich-Julian value
 would not be a bad approximation.

\subsubsection{Thickness of the outer gap}
Let's discuss now the thickness of the outer gap accelerator.
Zhang \& Cheng (2003) estimated
the  gap thickness  for the millisecond pulsars when the photon-photon
pair-creation process controls the gap activities.
 They have argued that the gap thickness is controlled
by the photon-photon  pair-creation process between
the $\gamma$-rays emitted in the
outer gap and the X-rays from the stellar surface, where the multiple
magnetic field dominates the global dipole field (Ruderman 1991, Arons 1993).
 They estimated the typical gap thickness divided by the light cylinder radius
at the light cylinder as
\begin{equation}
f_{ZC}=\frac{D_{\perp}(R_{lc})}{R_{lc}}=7.0\times 10^{-2}P_{-3}^{26/21}B_{8}^{-4/7}\delta r_{5}^{7/2}.
\label{fzc}
\end{equation}
Here $\delta r_5$ is the distance (in units of $10^{5}$~cm)
 from the stellar surface to the position where the local magnetic
field is equal to the dipole field. In the following we assume that $\delta r_5\sim 1-10$.

Takata et al. (2010a) proposed  the outer gap model, in which
 the gap thickness is determined by
 the magnetic pair-creation process near the stellar surface.
They argued that the returning particles, which were
accelerated  in the gap, will
emit photons with an energy  $m_ec^2/\alpha_f\sim 100 \rm
MeV$ by  curvature radiation near the stellar surface, where  $m_ec^2$
and $\alpha_f$ are the electron rest mass energy and the fine structure
constant, respectively.
The 100MeV photons form pairs by the magnetic
pair creation process. The photon multiplicity is easily
over $10^4$ for each incoming particle.  For a simple dipole field
structure, all  pairs should move inward and  cannot
affect the outer gap accelerator.
 However Takata et al. (2010a) have argued that  if the magnetic
field lines near the surface, instead of being
nearly perpendicular to the surface,  are bending side-ways due to the
strong local field, the pairs created on these local magnetic field
lines can have an angle bigger than 90$^{\circ}$, which results in an
outgoing flow of pairs.  With this model, the estimated
fractional gap thickness is given by
\begin{equation}
f_m=\frac{D_{\perp}(R_{lc})}{R_{lc}}\sim \frac{D_{\perp}(R_s)}{R_p}=
0.8 K  (P/1~\mathrm{s})^{1/2},
\label{fm}
\end{equation}
where $R_p\sim R_s(R_s/R_{lc})^{1/2}$ is the polar cap radius,
 $K\sim B_{m,12}^{-2}s_7$ is the parameter characterizing  the
local magnetic field properties, $B_{m,12}$ and $s_7$ are the local magnetic field
in units of $10^{12}$G and the local curvature radius in units
of $10^7$cm, respectively. By fitting the radiation properties of 
 the $\gamma$-ray pulsars observed  by $Fermi$, they find that $K\sim 2$ for the canonical pulsars, while $K\sim 15$ for the millisecond pulsars.
When   the fractional gap thickness $f_{m}$ is smaller  (or larger)
than  $f_{zc}$, the magnetic  pair-creation (or photon-photon pair-creation)
process controls  the gap thickness.

Finally we   discuss
the maximum fractional thickness, $f_{max}$, for
the active outer gap accelerator.
 Zhang \& Chang (1997) have argued that a pulsar with the fractional
gap thickness larger  than
unity, $f_{ZC}>1$, is not active, because the pairs are not created in the
gap by the photon-photon pair-creation process.  However, the
 outer gap accelerator can exist only between
the last-open field lines and the critical
field lines,  which are defined by those magnetic
field lines that have  the null point of the Goldreich-Julian 
charge density  at the light cylinder. Therefore,  we may define the
maximum gap thickness  as
\begin{equation}
f_{crit}=\frac{\theta_P-\theta_c}{\theta_p},
\end{equation}
where $\theta_p$ and $\theta_c$ are the polar angles of the last-closed 
field line and of the critical field line on the stellar surface,
respectively.  For the pure dipole field, we obtain
\begin{equation}
\theta_{p}=\alpha+\sin^{-1}\left[\sin(\theta_{lc}-\alpha)
\left(\frac{R_s}{R_{lc}}\sin\theta_{lc}\right)^{1/2}
\right]
\end{equation}
and
\begin{equation}
\theta_{c}=\alpha+\sin^{-1}\left[\sin(\theta_{n}-\alpha)
\left(\frac{R_s}{R_{lc}}\sin\theta_{n}\right)^{1/2}
\right ],
\end{equation}
respectively, where
\[
 \theta_{lc}=\tan^{-1}\left(\frac{-3-\sqrt{9+8\tan^2\alpha}}{4\tan\alpha}
\right),
\]
and
\[
\theta_n=\tan^{-1}\left(\frac{3\tan\alpha+\sqrt{8+9\tan^2\alpha}}{2}\right).
\]
In this paper,
we examine two extreme cases corresponding to $f_{max}$=1 and
 $f_{max}=f_{crit}$ respectively.
The results for these two cases may  give a range of  uncertainty
 of the present theoretical predictions.

\subsubsection{Dependence  on the inclination and viewing angles}
The observed characteristics of the $\gamma$-ray emissions
 depend on the viewing angle and the inclination angle between
the rotation axis and the magnetic axis.
With  the geometry of the outer gap model, stronger emissions are emitted toward
a viewing angle $\xi\sim 90^{\circ}$, and   the observed flux tends to
decrease as the  viewing angle closes to the rotation axis, where
$\xi=0^{\circ}$ or $\xi=180^{\circ}$. To take into account this effect, we
calculate the $\gamma$-ray flux, described in section~\ref{oflux},
as a function of the viewing angle and of the inclination angle.
In the calculation, we apply  the vacuum
dipole field geometry and assumed that
the outer gap  extends from the null charge point
of the Goldreich-Julian
charge density to $r=1.5R_{lc}$ or $\rho = R_{lc}$, where $\rho$ is the axial
distance from the rotation axis. We also assume  for simplicity no
 azimuthal dependence of the gap structure.

\begin{figure}
\begin{center}
\includegraphics{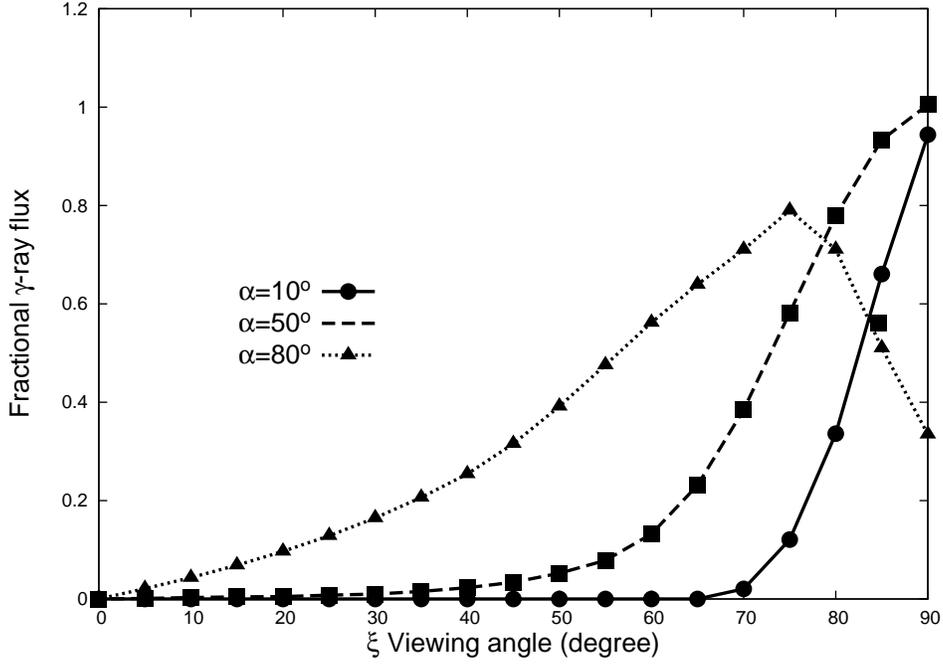}
\caption{The fractional $\gamma$-ray flux as a function of the viewing
angle and of the inclination angle.  The fractional flux is defined
 by the ratio of the calculated  flux measured on the Earth
to $f^3L_{sd}/d^2$. The results are for $P=3$~ms,
$B_s=3\times 10^{8}$~Gauss and $f(R_{lc})=0.4$. }
\label{gflux}
\end{center}
\end{figure}

Figure~\ref{gflux} shows the dependence of the $\gamma$-ray flux
as a function of the viewing angle ($\xi$)  and of the inclination
 angle ($\alpha$). The vertical axis represents the fractional $\gamma$-ray
flux, which is defined as the  $\gamma$-ray flux divided
by $f^3L_{sd}/d^2$.  The different lines correspond to the different
inclination angles.
The results are obtained for a rotation period
$P=3$~ms and the stellar magnetic field $B_s=3\times 10^{8}$~Gauss. We can see
 in Figure~\ref{gflux} that  the calculated flux tends to decrease as
the viewing angle\ and the inclination angles decreases from $90^{\circ}$.

We can see  that the general trend
 of the relation between  the fractional
$\gamma$-ray flux and the viewing angle and the inclination angle,  seen in
Figure~\ref{gflux},  is maintained for  different sets of  the rotation 
periods and  magnetic fields. In this study,  for simplicity, we use 
the relations in Figure~\ref{gflux} for all sets of the rotation period and 
of the magnetic field. We assume that the inclination angle and the viewing 
angle are randomly distributed.

 However,  we do not take into account the dependence
of the spin down power on
the inclination angle. The pulsar spin-down can result from both the current braking torque and
the magnetic dipole radiation. The former can spin-down the pulsar even if the inclination angle is zero.
According to the analysis of the force-free magnetosphere
done by Spitkovsky (2006), the spin down power depends
on the inclination angles as $L_{sd}\propto 1+\sin^2\alpha$; in other words,
$L_{sd}$  changes  only by a factor of 2  with the inclination angle.
 With this small variation, we may expect that
the general properties of the simulated distributions for
 the MSP's characteristics (such as the period and
the magnetic field) are not affected much by  dependence  on the
inclination angle.

\section{Results and Discussions}
\label{result}
In the Monte-Carlo simulations, we generated
 $3\times 10^6$ MSPs with a constant birth rate over 10~Gyr.
About 2100~radio MSPs were detected by the simulated radio surveys.
 Scaling the simulated
population to the number of  radio MSPs in the ATNF catalog, $\sim 80$,
we obtain  $\sim10^{-5}$ per year  as the
predicted averaged birth rate, which is consistent,
within the uncertainties, with the other population studies
(e.g. Ferrario \& Wickramasinghe 2007).

\subsection{Galactic distribution of MSPs}
\begin{figure}
\begin{center}
\includegraphics{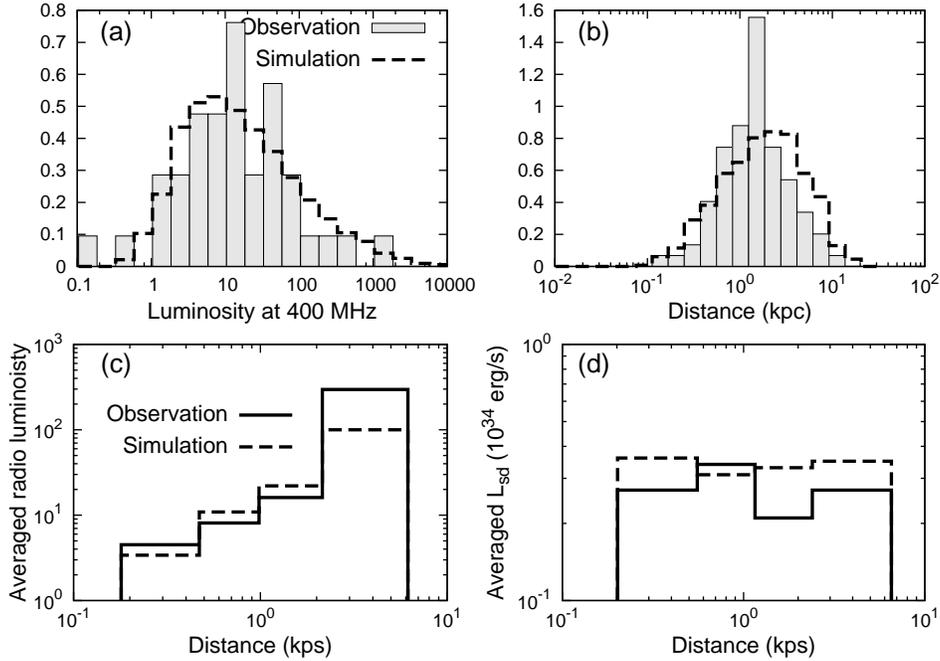}
\caption{Distributions of (a) luminosity  and  (b) distance
of the observed (shaded histograms) and simulated (dashed lines)
radio MSPs.  The panels (c) and (d) represent the averaged radio
luminosity and the spin down power  for the observed (solid lines)
and the simulated (dashed lines) MSPs
 as a function of the distance, respectively.
Each bin of the solid lines contains the same number of the observed MSPs.}
\label{rdis}
\end{center}
\end{figure}

In this section, we discuss the consistency of the observed and simulated
 distributions of the radio MSPs. First,  Figures~\ref{rdis}~(a) and~(b) compare
the observed (shaded histograms) and simulated (dashed lines)
 distributions of the radio luminosity
at 400~MHz and of   distance, respectively. We can see that  the
simulated distribution (dashed lines)
  qualitatively explain  the observed features; for example,
the  peak positions of the distributions. On the other hand,
we find that it is difficult
to explain the sharp peak at 1-2~kpc in the observed  distribution
of the distance. We argue that this
 excess in the observed distribution is caused by
several MSPs, and that the excess may be produced by
enhancements  of the local medium (e.g. the Gould Belt) and of the local
  birth rate.  For example, the age of the Gould Belt is believed to be 
 $\sim  10^{7-8}$~yr (Grenier 2004). Because
the birth rate of the canonical pulsars in the Gould Belt
is expected to be one per $10^{5}$ yr, and because the birth rates of
the MSPs in the Galactic field are 2-3 orders smaller than that of the canonical
pulsars,  one may expect  one per $10^{7-8}$~yr as
the birth rate of the MSPs in the Gould Belt. This indicates
 that several  MSPs can be born within  the age of the Gould Belt.
 If one assumes that the typical radial  velocity of MSPs
is $(70/\sqrt{3})$~km/s $\sim 40$ km/s, then  the displacement from the Gould Belt is
$d\sim 5\times 10^7~\mathrm{yr} \times  40~\mathrm{km/s}\sim 2$~kpc, which
is consistent with the enhancement of the peak in the distribution of the
distance.

Figures~\ref{rdis}(c) and (d) show  the averaged luminosity and
spin down power, respectively, as a function of the distance.
The solid and dashed lines are the results for the observations and for the
simulations, respectively, and each bin of the histograms has
the same number of observed MSPs. We find in Figures~\ref{rdis}(c) and (d)
 that the present simulation
describes characteristics of the observations very well, that is, the averaged
radio luminosity increases with the distance and the averaged spin down power
 is almost constant with the distance. Finally,  we can  see that
 the simulated MSPs can reproduce the Galactic longitude and latitude
distributions of  the radio MSPs. On these grounds,
 we conclude that our  simulation  reproduces  the
Galactic distribution  of the radio MSPs.

\subsection{Comparison with $Fermi$ observations}
\label{comparison}
In this section, we compare the results of our simulation with the $Fermi$ 
six-month long  observations. $Fermi$ has found 9 ``radio-selected'' $\gamma$-ray
emitting MSPs with a flux $F_{\gamma}\ge 10^{-11}~\mathrm{erg/cm^2 s}$
(Abdo et al 2009b; Abdo et al. 2010a,c), and all of them are located
 at high Galactic latitudes, that is, $|b|\ge 5^{\circ}$.  Abdo et al. (2010a)
 show that the sensitivity of the low Galactic latitudes is about a factor of
three worse as compared with that of high Galactic latitudes.
In the present simulation, therefore, we use
$F=10^{-11}~\mathrm{erg/cm^2 s}$ for the Galactic latitudes
$|b|\ge 5^{\circ}$ and $3\times 10^{-11}~\mathrm{erg/cm^2 s}$
for $|b|\le 5^{\circ}$ as the sensitivity requirement
 of the radio-selected MSPs for the six-month long observations. 
 Because no $\gamma$-ray-selected MSPs have been
detected  so far, we cannot simulate the  $Fermi$ sensitivity
of  the blind search for the MSPs.  In this section, we show the results of
our simulations by setting  the sensitivity to that of the canonical pulsars,
that is,  $F=2\times 10^{-11}~\mathrm{erg/cm^2 s}$ for  Galactic latitudes
$|b|\ge 5^{\circ}$ and $6\times 10^{-11}~\mathrm{erg/cm^2 s}$
for $|b|\le 5^{\circ}$.

Table~1 summarizes the population of the radio-selected
and the $\gamma$-ray-selected MSPs detected within the simulations.
$N_r$ and $N_g$ represent the population of the
  radio-selected and the $\gamma$-ray
selected MSPs, respectively,  detected with the $Fermi$ six-month  sensitivity
and  $N_{i,F>10^{-10}}$~($i=r,g$) are the populations of ``bright'' MSPs
with a $\gamma$-ray flux larger than $10^{-10}~\mathrm{erg/cm^2 s}$.
 In addition,
$N_{i,5}$ and $N_{i, 10}$ are the populations
with sensitivity  projected to 5~year and 10~year observations, respectively, 
for  which the sensitivity is scaled  $\propto \sqrt{T}$, 
where $T$ is the length of the observation time.

As Table~1 shows, our model predicts that only 2-3 radio-selected  and
1-3 $\gamma$-ray-selected  MSPs   can be
detected with a flux larger $10^{-10}~\mathrm{erg/cm^2 s}$.
 Although the results of the  $Fermi$ with the six-month  long  observations
(Abdo et al. 2009a) did  not include the bright sources,
it is  possible that with the small
population of the bright MSPs,
$Fermi$  may have  missed the detection of the pulsation. The
$Fermi$ unidentified catalog includes about 50 sources with a flux larger than
$10^{-10}~\mathrm{erg/cm^2 s}$. We can see that the spectral behaviors of 
most of them are consistent  with those of the known $\gamma$-ray pulsars;
 (1) they are steady sources in the sense that
the variability index defined in Abdo et al. (2010b) is smaller than 23,
  and (2) the spectra above 100~MeV cannot be fit well by single power law,
 that is, the curvature index, $C$  (defined in Abdo et al. 2010b),
is  $\ge 10$. Although most of them may be canonical $\gamma$-ray
pulsars (c.f. Takata et al. 2011) or Active Galactic Nuclei    (Abdo 2010c), it is very likely that
a small fraction, in particular, high Galactic sources,
 are $\gamma$-ray emitting MSPs (see section~\ref{gamsp}).
Recently, Ransom et al. (2011) reported the discovery of three radio MSPs
associated with the  $Fermi$ unidentified bright sources, suggesting 
 that our results are quite consistent with the observations.

Applying  the sensitivity of $Fermi$ six-month long observation, 
  our model predicts
9-13 radio-selected $\gamma$-ray MSPs, which is consistent with the 9 MSPs with 
 $Fermi$ observations. We also predict
 22-35 $\gamma$-ray-selected MSPs, whereas  $Fermi$
has had no such detections so far.
 For the canonical pulsars, the sensitivity of the blind search is about
a factor of 2 worse as compared with that of the radio-selected pulsars
(Abdo et al. 2010a). The population of the $\gamma$-ray-selected
pulsars is roughly comparable with that of the radio-selected
$\gamma$-ray pulsars.   On the other hand,
 it is expected that the detection  of the rotation period from
the MSPs (in particular in binary systems) by the blind search  is
even harder and the sensitivity  is
much worse than that of the canonical pulsars.

\begin{figure}
\begin{center}
\includegraphics{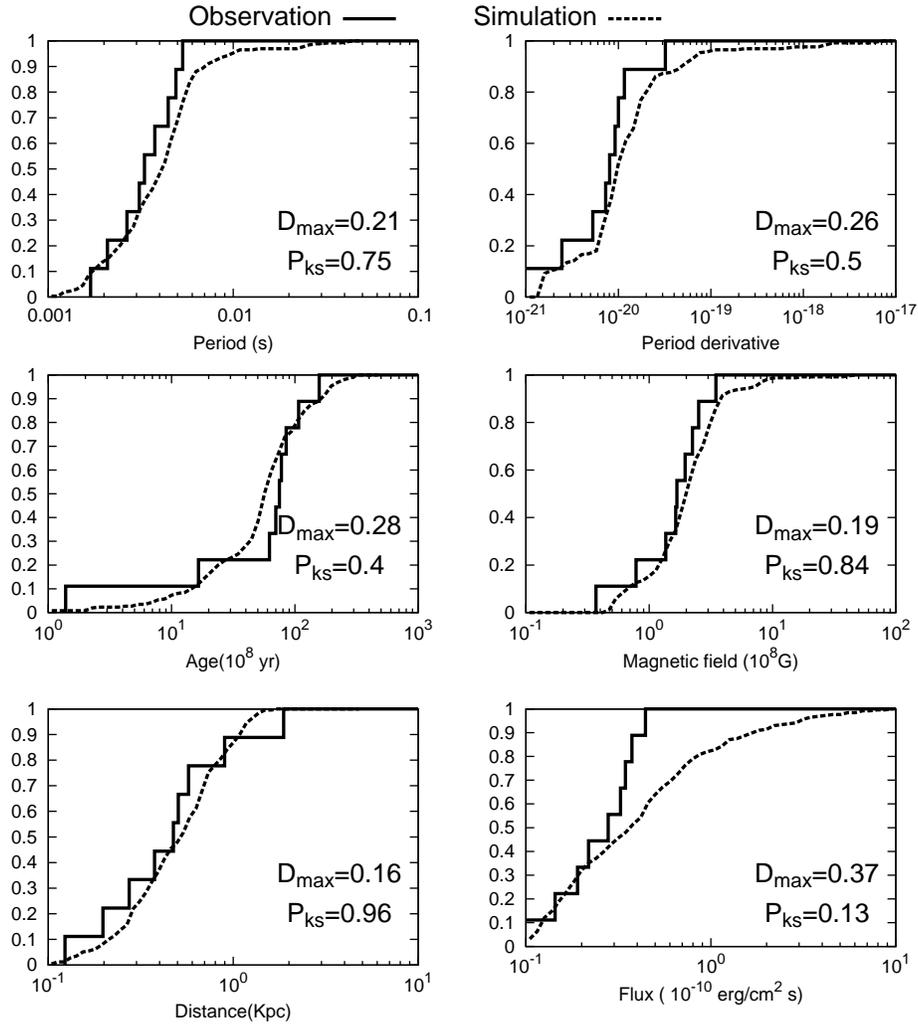}
\caption{Cumulative distributions for the various characteristics of
the observed (solid lines) and simulated (dashed lined)
 radio-selected $\gamma$-ray MSPs with the sensitivity of the 
 $Fermi$ six-month long observation. 
 The maximum difference ($D_{max}$)
and the p-value of KS-test ($P_{ks}$) are also displayed.
The intrinsic values, that is, the values without  the Shklovskii effect,
are also represented.
 }
\label{dispropr}
\end{center}
\end{figure}

Figure~\ref{dispropr} compares the cumulative  distributions of the
 various characteristics of
the observed  9 (solid lines)  $\gamma$-ray MSPs detected by
$Fermi$ (Abdo etal. 2009b; Abdo et al. 2010a,b,c)
 and the simulated (dashed lines) radio-selected  $\gamma$-ray MSPs.
The results are for $f_{max}=f_{crit}$.
We performed a Kolmogorov-Smirnov (KS) test to compare the two cumulative
distributions. In Figure~\ref{dispropr}, we present
 the maximum deviation ($D_{max}$) between the two
distributions and the p-value ($P_{ks}$) of the KS-test for each pulsar
characteristics. Since
the simulated sample has more than 400 pulsars, we used the one-sample 
KS statistic. For example, the p-value for the rotation period   is $P_{ks}\sim 0.75$, suggesting  that the hypothesis that the two distributions are 
drawn from the same distribution cannot be rejected at better than 25\% 
confidence level. We find in Figure~\ref{dispropr} that  for 
all distributions except that for the flux distribution, the hypothesis
 cannot be rejected at better than 60\%, indicating that the model
distributions are consistent with the observations.

For the $\gamma$-ray flux (right-bottom panel in Figure~\ref{dispropr}),
the result of the KS-statistic provides a relatively low p-value, as compared with
other characteristics.   This is because
all 9  MSPs detected by $Fermi$ have a $\gamma$-ray flux smaller than
 $10^{-10}~\mathrm{erg/cm^2 s}$, while the model predicts that about 20\%
of $\gamma$-ray MSPs  have a flux larger than
$10^{-10}~\mathrm{erg/cm^2 s}$.
However, we note that the difference between the
 observed  and  simulated distributions  is  caused by only one or two MSPs.
Hence, if several  MSPs with a bright $\gamma$-ray emissions of
$F_{\gamma}\ge 10^{-10}~\mathrm{erg/cm^2 s}$ have been missed by the
$Fermi$ observations, the model distribution is not in conflict
with the observations.  It is important to note that our simulation results are
based on the existing radio sensitivities of pulsar surveys. Any deep radio search implies
increasing the radio sensitivities and consequently some $gamma$-ray selected MSPs can become radio-loud.
In fact, several radio MSPs have been
detected  by deep search for
the  radio emissions from bright $Fermi$ unidentified sources
(Ransom et al. 2011). These radio-loud $gamma$-ray MSPs are defined as $gamma$-ray selected MSPs in our simulations.

As Table~1 shows, the model predicts that
 the population of radio-selected $\gamma$-ray MSPs increases by 
only about 10 over the 5-years (or 10-years) of  $Fermi$ observations.
 This implies that most of  the presently
known radio MSPs ($\sim 80$) might not be discovered by $Fermi$. 
However, the predicted radio-loud $\gamma$-ray MSPs depend on 
the sensitivity of the radio surveys, as we can see in Table~2. 
For the $\gamma$-ray selected MSPs,  the simulation predicts that $Fermi$ can 
detect at least 100 sources. 
 As we will discuss in section~\ref{gamsp}, we
expect that these simulated $\gamma$-ray selected MSPs
correspond to  the $Fermi$ unidentified sources. As we have emphasized before, 
we predict that more and more  radio
MSPs will be confirmed among the $Fermi$ unidentified sources
by a deep search of the radio emissions.

Story et al. (2007) studied the population of $\gamma$-ray MSPs by using
the slot gap accelerator model and predicted that $Fermi$ would 
 detect about 12 radio-loud
and 33-40 radio-quiet $\gamma$-ray MSPs. On the other hand, our outer gap model
tends to predict more $\gamma$-ray (in particular  radio-quiet)
MSPs than the slot gap model. Also,
the present simulation  predicts a larger
 ratio between the radio-quiet and radio-loud
 MSPs  ($\sim 6-7$) than that predicted by the slot gap model
($\sim 3$), although the ratio depends on the simulated sensitivities of
 the $Fermi$ and of the radio observations (sections~3.3 and~3.4).

\subsection{Population of $\gamma$-ray MSPs}
\label{population}
\begin{figure}
\begin{center}
\includegraphics{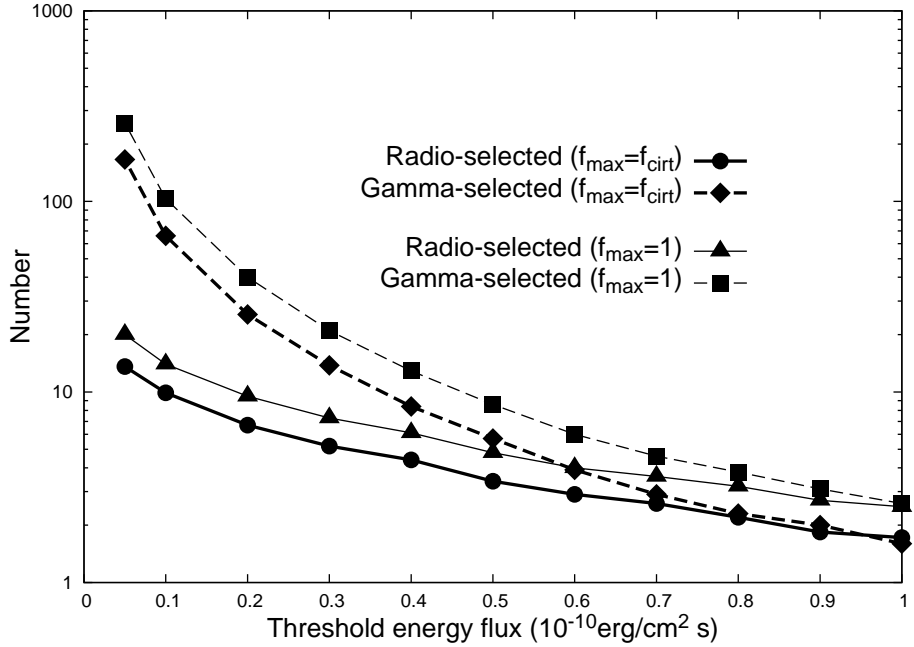}
\caption{The simulated numbers of the radio-selected (solid line)
and $\gamma$-ray-selected  (dashed line) $\gamma$-ray MSPs as a function
of the threshold energy flux.   }
\label{number}
\end{center}
\end{figure}

In Figure~\ref{number}, we summarize the population of the $\gamma$-ray emitting
 MSPs detected  within the simulation as a function of
the threshold energy flux of the $\gamma$-rays. The solid
lines and dashed lines represent the  population  of the radio-selected and
 $\gamma$-ray-selected $\gamma$-ray MSPs respectively. The thick and thin
 lines represent results for the maximum fractional gap thickness of
$f_{max}=f_{crit}$ and $f_{max}=1$, respectively. For example,
the present simulation predicts
that $\sim 2-3$ and  $\sim 10-14$ radio-selected
 $\gamma$-ray MSPs irradiate
the Earth with a $\gamma$-ray flux $F_{\gamma}\ge 10^{-10}$ and
$F_{\gamma}\ge 10^{-11}~\mathrm{erg/cm^2 s}$, respectively.
 For the radio-quiet $\gamma$-ray MSPs, about 100 sources 
with a flux larger than $10^{-11}~\mathrm{erg/cm^2 s}$ irradiate  the Earth

We can see in Figure~\ref{number} that the numbers of $\gamma$-ray-selected MSPs increase
more  rapidly than that of the radio-selected $\gamma$-ray MSPs  and the
ratio of  the $\gamma$-ray-selected to radio-selected $\gamma$-ray MSPs
increase with decreasing the threshold energy flux.
As the threshold of $\gamma$-ray energy flux  decreases,
 the $\gamma$-ray emissions from
more distant MSPs can be detected in the simulations.
On the other hand, because we fix the sensitivity of the radio surveys in
Figure~\ref{number},  the radio emissions from those  distant  MSPs  may  
not be detected by the radio surveys. As a result, more $\gamma$-ray selected 
MSPs are  detected in the simulations as the threshold flux decreases. In fact, if we count all
$\gamma$-ray MSPs irradiating the Earth with the radio emissions as the
radio-selected pulsars, the ratio does not depend on the
 threshold energy flux of the $\gamma$-ray emissions.

\begin{figure}
\begin{center}
\includegraphics{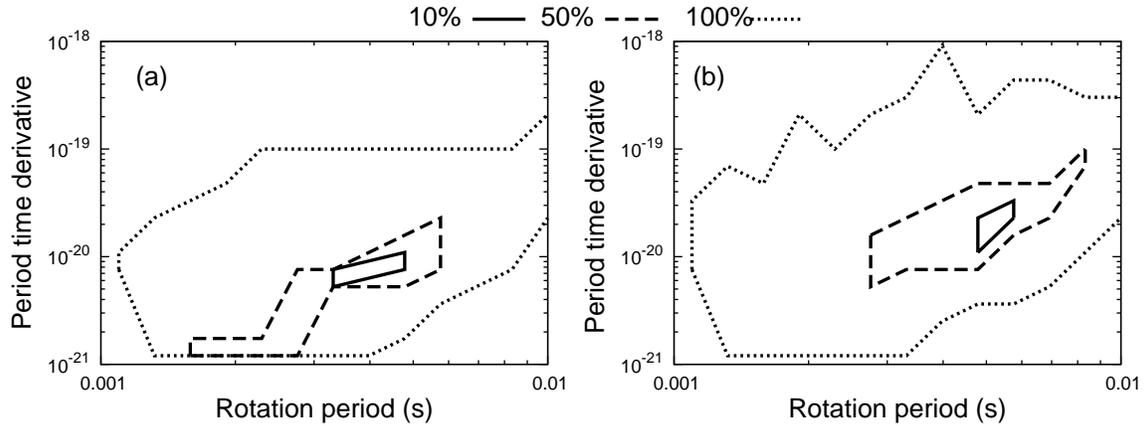}
\caption{Contours of two-dimensional (rotation period and period time
derivative)
distributions of the simulated MSPs with the sensitivity of the 
$Fermi$ six-month long observation;
(a) intrinsic period time derivative
and (b) ``observed'' period time derivative adding   Shklovskii effect.
More MSPs are located within the solid lines.
  Within solid, dashed and dotted  lines,
  10\%, 50\% and 100\% of the total MSPs are populated. }
\label{dis2D}
\end{center}
\end{figure}

Figure~\ref{dis2D}  shows  contours of  the two-dimensional distribution
in the rotation period and the period-time derivative of the simulated
 MSPs  with a $\gamma$-ray flux
 $F_{\gamma}\ge 10^{-11}~\mathrm{erg/cm^2 s}$. We present  the distributions
for all simulated MSPs including the radio-selected and $\gamma$-ray-selected
pulsars, because the  two distributions do not  differ  much from each other.
The  left panel shows the distributions using
 the period time derivative for the intrinsic value,
 and the right panel represents the period time derivative
for  the ``observed'' value, that is, the value  after adding
the  Shklovskii effect described by equation~(\ref{shk}).
Within solid, dashed and dotted  lines,
  10\%, 50\% and 100\% of the total MSPs are populated.
  We can see that the Shklovskii effect slightly  shifts  the peak of the distributions toward longer
 rotation periods and larger period time derivatives.  For the observational
distributions (right panel), about 50\% of the
$\gamma$-ray emitting MSPs will be detected with a    rotation period
between $P\sim 0.003-0.008$~s and a  period time derivative between
$\dot{P}\sim 6\times 10^{-21}-10^{-19}$. This information can be used
to narrow  down the parameter range for a blind search to detect
 the rotation period from $Fermi$ unidentified sources.

\subsection{Origin of the  Galactic $Fermi$ unidentified steady sources}

\label{gamsp}
As Figure~\ref{number} shows, our simulation predicts that about $\sim 200$
 $\gamma$-ray MSPs irradiate
 the Earth with a flux $F_{\gamma}\ge 5\times 10^{-12}~\mathrm{erg/cm^2 s}$
 and with the present sensitivity of the radio surveys most of them are categorized as $\gamma$-ray-selected MSPs.
Therefore,  it is likely that although 
 millisecond pulsed emission has not been
confirmed yet, the $\gamma$-ray MSPs contribute to the galactic
unidentified $Fermi$ sources, such as the  newly discovered
20~radio MSPs associated with the $Fermi$ unidentified sources (Ray 2010;
Caraveo 2010; Ransom et al. 2011).

\begin{figure}
\begin{center}
\includegraphics{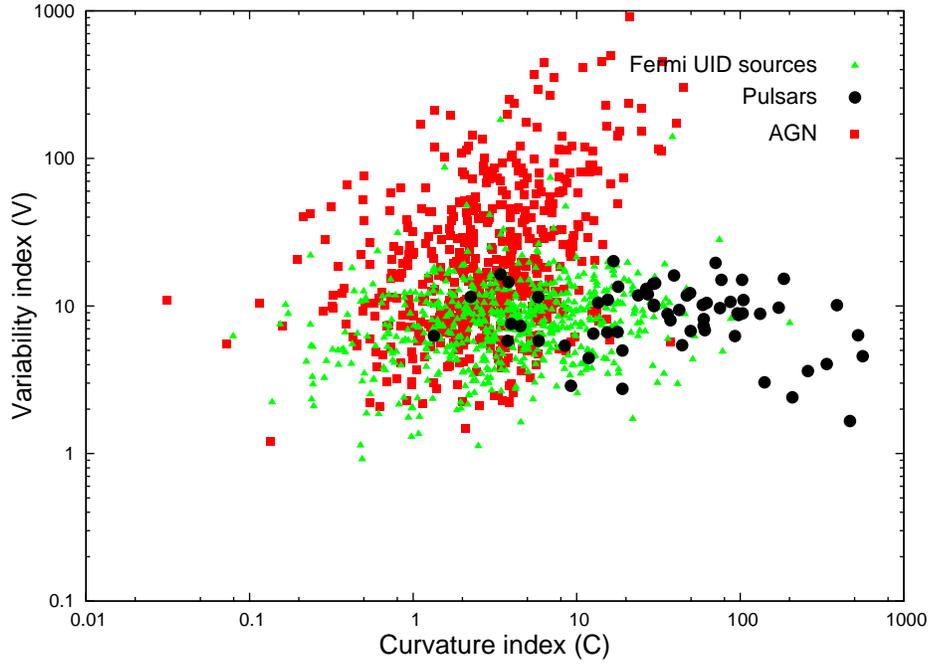}
\caption{Plots for the curvature index and the variability index of
the $Fermi$ sources (Abdo et al., 2010b). The filled circles, and boxes
and triangles are the pulsars, AGNs, and the $Fermi$ unidentified sources respectively. }
\label{vc1}
\end{center}
\end{figure}

Figure~\ref{vc1} plots the curvature index (C) and the variability index (V)
of the  $Fermi$ sources; a curvature index larger than
11.34 indicates a less than 1\% chance
that the power-law spectrum is a good fit above 100~MeV, and   a variability
 index larger than 23.21 indicates less than a 1\% chance of being a steady
source (see  Abdo et al. 2010b for the exact definition of C and V indexes).
 The filled circles, boxes and triangles correspond to
the pulsars, Active Galactic Nuclei (AGN) and $Fermi$ unidentified sources.
 In Figure~\ref{vc1}, we can easily see that the pulsars and the
 AGN belong to different groups
 in the C-V plane, that is, the pulsars belong to $C\ge 5$ and $V\le 20$,
whereas  AGN belong to wide range of variability  index, which
may indicate a correlation between the C-index and V-index (that is,
 $C\propto V$).  Most of the $Fermi$ unidentified sources have
the V-index smaller than $\sim$30, with the C-index larger than 0.1.
Figure~\ref{vc1} indicates that the $Fermi$ unidentified sources can be
either pulsars or AGNs.

\begin{figure}
\begin{center}
\includegraphics{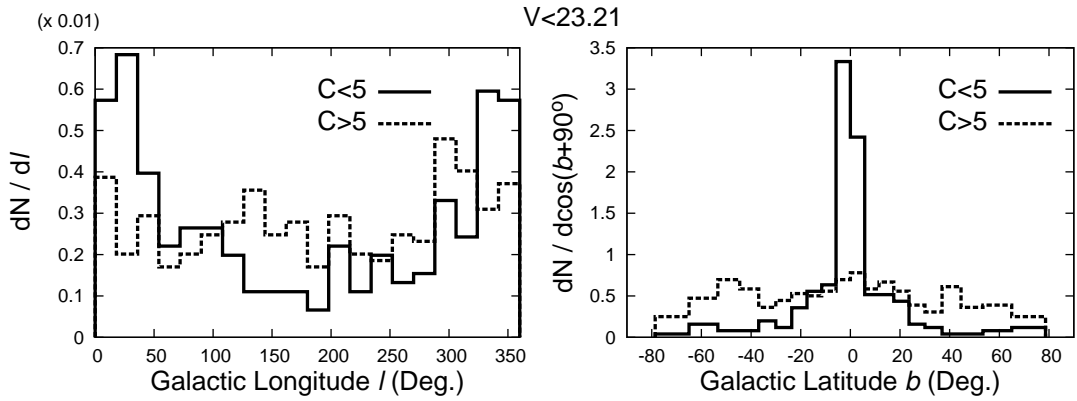}
\caption{The distributions in  Galactic coordinate longitudes (left) and
latitudes (right) of the
$Fermi$ unidentified steady sources.  The solid and dashed lines are the 
results for $C<5$ and $C\ge 5$, respectively. }
\label{galc}
\end{center}
\end{figure}

In Figure~\ref{galc}, we plot the Galactic distributions of
the $Fermi$  unidentified steady sources for $C<5$ (solid line) and $C\ge 5$
(dashed line).   It is clear from Figure~\ref{galc} that the two
lines represent  different distributions with respect to each other. The solid lines show
relatively  constant distributions for the Galactic longitudes and  latitudes,
indicating that most of the $Fermi$ unidentified sources with $C<5$ (dashed lines)
distribute isotropically in the sky, and may be related with
extra-Galactic  (or Galactic halo) sources.  For the dashed-lines,
 on the other hand, the distributions for the Galactic longitudes and latitudes
coordinate show  peaks at the Galactic centre and at the Galactic disk,
 respectively, suggesting  that the $Fermi$ unidentified steady sources with
$C\ge 5$ are associated with Galactic objects.

\begin{figure}
\begin{center}
\includegraphics{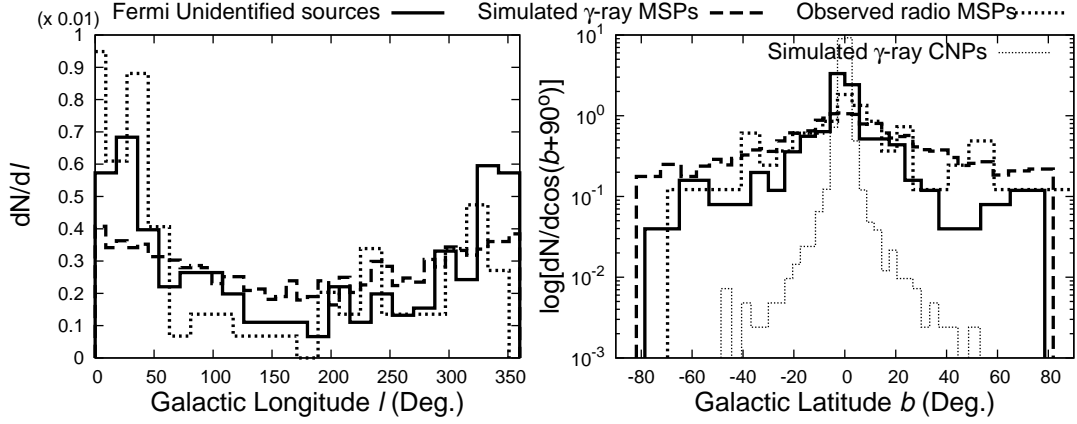}
\caption{The distributions of  Galactic longitudes (left)
and latitudes (right) of the $Fermi$ unidentified point sources with
$V\le 23.21$ and $C \ge  5$ (solid lines),
the simulated $\gamma$-ray
MSPs (dashed lines) and observed radio MSPs (dotted lines).  The distribution
of  Galactic latitudes of the simulated $\gamma$-ray emitting canonical
pulsars (CNP) in Takata et al. (2010b) are also plotted in the right
 panel (thin dotted line).}
\label{galMSP}
\end{center}
\end{figure}

In Figure~\ref{galMSP}, we compare
the distributions of the Galactic longitudes (left panel) and latitudes
(right panel) for the unidentified $Fermi$
sources with $V\le 23.21$ and $C\ge 5$ (solid line) and the
simulated $\gamma$-ray MSPs (dashed line),
 with a flux $F_{\gamma}\ge\ 10^{-11}~\mathrm{erg/cm^2 s}$.
 The latitude distribution
of the $\gamma$-ray emitting canonical
 pulsars (dotted line) simulated in Takata et al. (2011)
and  the Galactic distributions for known radio MSPs (dotted lines)
are also plotted in the figure.

In Figures~\ref{galMSP}  we can see that the distributions of
the simulated $\gamma$-ray MSPs and of the observed radio MSPs are  consistent
 with that of the $Fermi$ unidentified sources, that is, (1)
three  longitude distributions have a peak  around the direction of the
Galactic centre $l=0^{\circ}$ and become  minimum around $l\sim 180^{\circ}$,
 and (2) the latitude  distributions have a peak at the Galactic
plane ($b=0$) and then they decreas
with the increase of the Galactic latitudes.  In particular, 
the MSPs can explain the distributions of the $Fermi$ unidentified sources 
located above the Galactic plane $|b|\ge 10$, which cannot be explained by the
canonical $\gamma$-ray  pulsars, as the right panel in Figure~\ref{galMSP} shows;
 the $\gamma$-ray emitting canonical pulsars can mainly explain the
unidentified sources located in the Galactic plane.
Since the MSPs are in general older
than  the canonical  pulsars, a higher
fraction of the $\gamma$-ray MSPs, as compared with the
canonical pulsars,  is located at
  higher Galactic latitudes. On these ground,  we conclude
 that $\gamma$-ray emitting MSPs are more plausible as 
candidates for the origin of the majority of the Galactic $Fermi$ unidentified
steady sources located in  high Galactic latitudes.

We note that  new radio MSPs have been discovered  in
the direction of about 20 unidentified $Fermi$ sources by the radio searches
 for individual source (e.g. Ray 2010; Caraveo 2010; Ransom et al. 2011).
These {\it $\gamma$-ray-selected radio-loud}  MSPs can 
  be classified  as the $\gamma$-ray-selected radio-quiet  sources
in the present simulation, because we do not take into account
 the radio search  for the specific sources.
In Table~2, on the other hand, we present how the population of $\gamma$-ray
emitting MSPs depends on the sensitivity of the radio survey. 
We used the Arecibo~2 (A2) and Parks~2 (P2) surveys (the first row),
 all radio surveys (the second  row) listed in table~1 of Takata et al. (2011)
 and  all radio survey but we
increase by a factor of 2  the sensitivity of each survey (third row).
The intrinsic populations, that is, the populations associated
with only beaming effects of the radio emission are shown in the bottom row.

We can see in Table~2 that the number of radio-selected pulsars increases with
the increase of the sensitivity of the radio surveys, whereas  the
$\gamma$-ray-selected pulsars decrease. This is because  if we increase the  sensitivity of the radio surveys some
 $\gamma$-ray-selected MSPs irradiating the Earth with  the radio emissions
may  be  re-classified as  radio-selected $\gamma$-ray pulsars.
As the bottom row in Table~2
shows, the intrinsic ratio of the  radio-loud and radio-quiet $\gamma$-ray
MSPs is 10$\sim 20$, indicating that most of
the $\gamma$-ray emitting  MSPs   irradiate the Earth with a corresponding 
radio emission. Therefore, our prediction is that as the sensitivity
of the radio observations will improve in the future, more  and more
radio MSPs will be discovered among 
 the $Fermi$ unidentified sources.

\begin{figure}
\begin{center}
\includegraphics{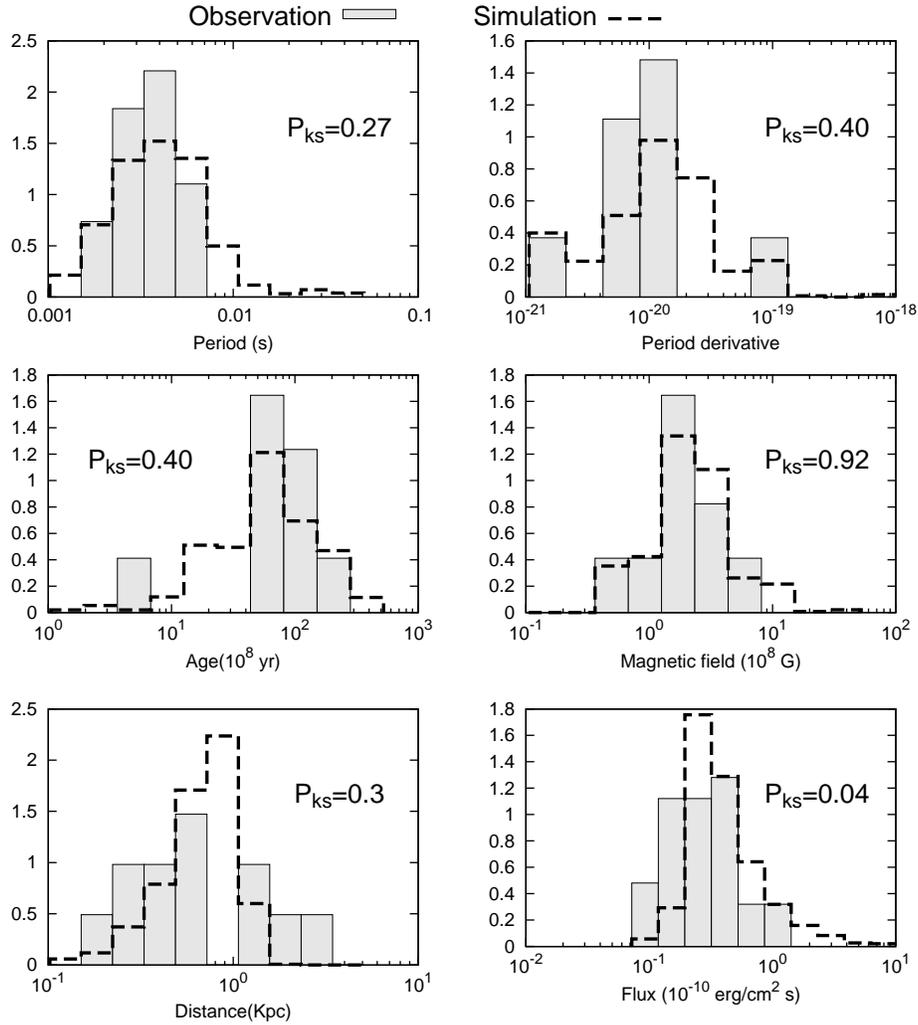}
\caption{Distributions of the various properties of the observed
(shaded histograms) and of the simulated (dashed lines) MSPs.
In addition to the 9 known $\gamma$-ray MSPs, the observed distributions of
the rotation period and of the distance includes 7 and 3 radio MSPs,
respectively, associated with the $Fermi$ unidentified sources.
For the $\gamma$-ray flux, the observed distributions includes 20 radio MSPs
 associated with the $Fermi$ unidentified sources. For the simulation,
both radio-selected and $\gamma$-ray-selected MSPs are included in the
distributions.}
\label{disprop}
\end{center}
\end{figure}

In Figure~\ref{disprop}, we present the distributions of various
characteristics of observed $\gamma$-ray MSPs (shaded histograms),
including the 20 new radio MSPs associated with  $Fermi$ unidentified
 sources,  and the simulated  $\gamma$-ray MSPs (dashed lines) with
the  simulated sensitivity of the $Fermi$ six-month long observation.
For the simulated distribution, both
the radio-selected and $\gamma$-ray-selected MSPs are taken into account.
 For the observations,
we were able to obtain   information of the rotation period for 7 out
the 20 radio  MSPs and
the distances for 3 MSPs (Ray, 2010; Ransom et al. 2011).  Therefore
 the distributions of the rotation period and of the distance
in Figure~\ref{disprop} are obtained from  16
(9 known $Fermi$ MSPs plus 7 radio MSPs) and 12 samples, respectively.
 The distributions of the period time derivative, the magnetic field
 and the spin  down age include only 9 known $Fermi$ MSPs,
because we could not find any
references for the intrinsic period time derivative of the 20~radio MSPs.
The $\gamma$-ray fluxes of the 20~radio MSPs are taken from
the $Fermi$ first catalog
(Abdo et al. 2010b~\footnote{see also
$\mathrm{http://fermi.gsfc.nasa.gov/ssc/data/access/lat/1yr_{-}catalog}$})
and Ransom et al. (2011).  In Figure~\ref{disprop}, we also indicate
the p-values  ($P_{ks}$)
 of the KS test.  As Figure~\ref{disprop}
shows, the simulated distributions are qualitatively consistent with the
observations (e.g. the position of the peak of the distributions).
 For the $\gamma$-ray flux, however, the p-value of KS-test is much lower than
those of other characteristics, indicating that the two distributions cannot be
drawn from same distribution. Because it is  expected  that the $\gamma$-ray
 emissions from the 20~radio MSPs associated with $Fermi$ unidentified
sources will be composed of a pulsed and an unpulsed component,  the
extraction of the pulsed component, that is,
 the detection of the pulsed period with $Fermi$,  is required
 to constrain our model.

We have assumed that MSPs are activated as the rotation powered pulsar as 
a result of the accretion process from the companion star. 
Recently Takata, Cheng \& Taam (2010b) argued that the
outer gap activities of MSPs can be turned on in the quiescent state of low
mass binary systems, for example PSR J1023+0038 (Archibald et al. 2009). The
geometry of the outer gap can direct the $gamma$-rays emitted from the gap
towards the companion star and the accretion disk. Consequently, the accretion
disk can be evaporated, and  the companion star becomes brighter due to the
irradiation of $gamma$-rays. If the high latitude unidentified sources of
$Fermi$ are indeed MSPs, some of them may be low mass binaries in a quiescent
state. We expect them to be associated with abnormally bright stars with
short orbital period. By searching the orbital periods of the optical
companion stars, we may be able to obtain the spin periods of the
radio-quiet MSPs from $gamma$-ray data.

\section{Conclusion}
\label{conclusion}
We have studied the population of the $\gamma$-ray emitting
millisecond pulsars using Monte-Carlo techniques.
We have applied the outer gap model with a switching of the gap closure
process from the photon-photon pair-creation model to the magnetic
pair-creation model, as suggested by Takata et al. (2010a).
Using the  sensitivity of the $Fermi$ six-month long observations,
 9-13 radio-selected $\gamma$-ray MSPs are detected within the simulation,
which is  consistent with the present 9 $Fermi$ MSPs.
The simulated  distributions for the various  characteristic of
 the radio-selected $\gamma$-ray pulsars
 are consistent with the present $Fermi$ observations
(Figure~\ref{dispropr}).
Scaling the observed sensitivity  $\propto \sqrt{T}$, the present model
predicts that $Fermi$  should detect 15-22
radio-selected $\gamma$-ray MSPs  and 95-152 $\gamma$-ray-selected  MSPs
 over its 5-year mission (Table~1).

 Our simulation predicts that about 100 (or 200-300)
 $\gamma$-ray  MSPs irradiate the Earth with a flux 
 $F\ge 10^{-11}~\mathrm{erg/cm^2 s}$
(or $5\times10^{-12}~\mathrm{erg/cm^2 s}$), and most of them are categorized as
  $\gamma$-ray-selected pulsars with the present sensitivity 
of the radio surveys (Figure~\ref{number}).
Our simulation also predicts that about 50~\% of
the $\gamma$-ray emitting MSPs will be detected with a rotation period
 in the range $P\sim 3-8$~ms and a period time derivative in the range 
$\dot{P}\sim 6\times 10^{-21}-10^{-19}$.  We further argue that
 $\gamma$-ray emitting  MSPs are plausible as candidates for
 the Galactic $Fermi$ unidentified steady
sources, located in  high Galactic latitudes.
 Our simulation implies that some of the  radio-quiet $\gamma$-ray
MSPs can be changed to radio-loud  $\gamma$-ray MSPs
as the simulated sensitivity of the radio surveys increasees.
Therefore, we predict that more and more
radio  MSPs will be discovered in
 the $Fermi$ unidentified sources as the radio sensitivity is improved.

\section*{Acknowledgement}
 We thank A.H. Kong, C.Y.~Hui, B.~Rudak,
M.Ruderman, R.E. Taam and  S.Shibata for the useful discussions,
 and T. Harko  and K. MacKeown for a critical reading of our manuscript.
We express our appreciation to an anonymous referee for  insightful
 comments.  We also thank the Theoretical Institute
for Advanced Research in Astrophysics (TIARA) operated under the Academia
Sinica Institute of Astronomy and Astrophysics, Taiwan,
which  enabled author (J.T.) to use the PC cluster at TIARA.
KSC is supported by a 2011 GRF grant of the Hong Kong SAR
Government entitled "Gamma-ray Pulsars".

\newpage

\begin{table}
\begin{tabular}{c|cc|cc|cc|cc}
\multicolumn{1}{c}{} & \multicolumn{1}{c}{$N_{r,F\ge 10^{-10}}$} & $N_{g,F\ge 10
^{-10}}$ & $N_{r}$ & $N_{g}$ & $N_{r,5}$ & $N_{g,5}$ & $N_{r,10}$ & $N_{g,10}$\\
\hline\hline
$f_{max}=f_{crit}$ & 2 & 2 & 9 & 22 & 15 & 101 & 17 & 157\\
$f_{max}=1$ & 3 & 3 & 13 & 35 & 22 & 161 & 25 & 248\\
\hline
\end{tabular}
\caption{Population of the simulated radio-selected  and $\gamma$-ray-selected
$\gamma$-ray MSPs. The subscript $r$ and  $g$ represent the radio-selected and
$\gamma$-ray-selected pulsars, respectively. $N_{i,F>10^{-10}}$~($i=r,g$) are
population of ``bright'' MSPs with a $\gamma$-ray flux larger
than $10^{-10}~\mathrm{erg/cm^2 s}$, and $N_{i}$ are results with the $Fermi$
six-month long observation.  In addition, $N_{i, 5}$ and $N_{i, 10}$
 are population with the sensitivity projected to 5~yr and 10~yr observations, respectively. }
\end{table}

\begin{table}
\begin{tabular}{ccccc|cccc|cccc}& \multicolumn{4}{c|}{6-month} & \multicolumn{4}{c|}{5-year} & \multicolumn{4}{c}{10-year} \\
 & \multicolumn{2}{c}{$f_{max}=f_{crit}$} & \multicolumn{2}{c|}{$f_{max}=1$} & \multicolumn{2}{c}{$f_{max}=f_{crit}$} & \multicolumn{2}{c|}{$f_{max}=1$} & \multicolumn{2}{c}{$f_{max}=f_{crit}$} & \multicolumn{2}{c}{$f_{max}=1$} \\
 & $N_r$ & $N_g$ & $N_r$ & $N_g$ & $N_{r,5}$ & $N_{g,5}$ & $N_{r,5}$ & $N_{g,5}$
& $N_{r,10}$ & $N_{g,10}$ & $N_{r,10}$ & $N_{g,10}$ \\
\hline\hline
A2, P2 & 5 & 25 & 7 & 40 & 9 & 105 & 12 & 168 & 10 & 162 & 14 & 255 \\
All & 9 & 22 & 13 & 35 & 15 & 101 & 22 & 161 & 17 & 157 & 25 & 248 \\
All ($\times 2$) & 16 & 18 & 23 & 30 & 26 & 93 & 37 & 150 & 29 & 148 & 42 & 234
\\
Intrinsic & 61 & 3 & 88 & 6 & 226 & 10 & 329 & 27 & 331 & 16 & 484 & 41 \\
\hline
\end{tabular}
\caption{Population of the simulated radio-selected and $\gamma$-ray-selected
 MSPs for the different radio surveys and projected mission lengths of $Fermi$(6 months, 5 years and 10 years). A2 and P2 represent
 the Arecibo~2  and Parks~2 surveys. ``All'' represents all
radio surveys  listed in table~1 of Takata et al. (2010b)
 and  ``All$\times 2$'' shows the results with  all radio survey but we
increase a factor of 2 for the sensitivity of each survey.
In addition, the intrinsic populations, that is, the populations associated
with only beaming effects of the radio emission,
 are shown in the bottom line. In this extreme limit, most $Fermi$ $\gamma$-ray pulsars become radio-selected $\gamma$-ray pulsars.}
\end{table}

\newpage

\newpage

\newpage

\label{lastpage}


\begin{thebibliography}{}
\bibitem[\protect\citeauthoryear{Abdo}{2010}]{ab010a}
Abdo A.A. et al., 2010a, ApJS,  187, 460
\bibitem[\protect\citeauthoryear{Abdo}{2010}]{ab010b}
Abdo A.A. et al., 2010b, ApJS,  188, 405
\bibitem[\protect\citeauthoryear{Abdo}{2010}]{ab010c}
Abdo A.A. et al., 2010c, ApJ, 712, 957
\bibitem[\protect\citeauthoryear{Abdo}{2010}]{ab010d}
Abdo A.A. et al., 2010d, ApJ, 715, 429
\bibitem[\protect\citeauthoryear{Abdo}{2009}]{ab09a}
Abdo A.A. et al., 2009a, Sci., 325, 840
\bibitem[\protect\citeauthoryear{Abdo}{2009}]{ab09b}
Abdo A.A. et al., 2009b, Sci., 325, 848
\bibitem[\protect\citeauthoryear{Abdo}{2009}]{ab09c}
Abdo A.A. et al., 2009c, ApJ, 706, 1331
\bibitem[\protect\citeauthoryear{Aliu}{2008}]{al08}
Aliu, E. et al., 2008, Sci, 322, 1221
\bibitem[\protect\citeauthoryear{Archibald}{2009}]{ar09}
Archibald, et al. 2009, Sci, 324, 1411
\bibitem[\protect\citeauthoryear{Arons}{1993}]{ar93}
Arons J., 1993, ApJ, 408, 160
\bibitem[\protect\citeauthoryear{Arons}{1983}]{ar83}
Arons J., 1983, ApJ, 266, 215
\bibitem[\protect\citeauthoryear{Binney}{1987}]{bi87}
Binney J. J., Tremaine S. D., 1987, Galactic Dynamics. Princeton Univ.
   Press, Princeton
\bibitem[\protect\citeauthoryear{burton}{1978}]{bu78}
Burton, W.B. \&  Gordon, M.A., 1978, A\&A, 63, 7
\bibitem[\protect\citeauthoryear{Campana}{1998}]{co98}
 Campana, S., Colpi, M., Mereghetti, S., Stella, L., Tavani, M., 1998, A\&ARv,
8, 279
\bibitem[\protect\citeauthoryear{Caraveo}{2010}]{ca10}
Caraveo, P.A., 2010, in High Time REsolution Astrophysics IV-The Era of Extremely Large Telescopes-HTRA-IV, Creece Ma7 5-7, arXIV:1009.2421

\bibitem[\protect\citeauthoryear{Cheng}{2000}]{ch00}
Cheng K.S., Ruderman M. \& Zhang L. 2000, ApJ, 537, 964
\bibitem[\protect\citeauthoryear{Cheng}{1998}]{ch98}
 Cheng K.S. \& Zhang, L. 1998, ApJ, 498, 327
\bibitem[\protect\citeauthoryear{Cheng}{1986a}]{ch86a}
 Cheng K.S., Ho C.,  Ruderman M. 1986a, ApJ, 300, 500
\bibitem[\protect\citeauthoryear{Cheng}{1986b}]{ch86b}
 Cheng K.S., Ho C., Ruderman M. 1986b, ApJ, 300, 522
\bibitem[\protect\citeauthoryear{Cordes}{2002}]{co02}
 Cordes, J.M. \& Lazio, T.J.W., 2002, preprint (astro-ph/0207156)
\bibitem[\protect\citeauthoryear{Daugherty}{1996}]{da96}
Daugherty J.K. \&  Harding, A.K., 1996, ApJ, 458, 278
\bibitem[\protect\citeauthoryear{Daugherty}{1982}]{da82}
Daugherty J.K., Harding, A.K., 1982, ApJ, 252, 337
\bibitem[\protect\citeauthoryear{Emmering}{1989}]{em89}
Emmering, R.T. \&  Chevalier, R.A., 1989, ApJ, 345, 931
\bibitem[\protect\citeauthoryear{Ferrario}{2007}]{fe07}
Ferrario, L., Wickramasinghe, D., 2007, MNRAS, 375, 1009
\bibitem[\protect\citeauthoryear{Frank}{2002}]{fa02}
Frank J., King A., Raine D., 2002, Accretion Power in Astrophysics, 3rd
   edn. Cambridge Univ. Press, Cambridge
\bibitem[\protect\citeauthoryear{Goldreich}{1992}]{go92}
Goldreich P.,  Reisenegger  A. 1992, ApJ, 395, 250
\bibitem[\protect\citeauthoryear{Goldreich}{1969}]{go69}
Goldreich P.,   Julian W.H. 1969, ApJ, 157, 869
\bibitem[\protect\citeauthoryear{Gonthier}{2002}]{Go02}
Gonthier, P.L., Ouellette, M.S., Berrier, J., O'Brien, S, Harding, A.K., 2002,
ApJ, 565, 482
\bibitem[\protect\citeauthoryear{Grenier}{2002}]{gr02}
Grenier I.A., 2004, preprint (astro-ph/0409096)
\bibitem[\protect\citeauthoryear{Hirotani}{2008}]{hi08}
Hirotani K., 2008, ApJL, 688, 25
\bibitem[\protect\citeauthoryear{Jeffrey}{1986}]{je86}
Jeffrey, L.C., 1986, Nature, 319, 384
\bibitem[\protect\citeauthoryear{Kramer}{1998}]{kr98}
Kramer, et al. 1997 A\&A, 322, 846
\bibitem[\protect\citeauthoryear{Kramer}{2000}]{kr00}
Kramer, M., Xilouris, K.M., 2000, in Pulsar Astronomy – 2000 and
  Beyond, ed. M. Kramer, N. Wex, \& R. Wielebinski (San Francisco:
  Astronomical Society of the Pacific), IAU Coll., 177, 229
\bibitem[\protect\citeauthoryear{Lorimer}{2008}]{lo08}
Lorimer D.R., 2008 Living Rev. Relaiv., 11, 8
\bibitem[\protect\citeauthoryear{Lorimer}{1995}]{lo95}
Lorimer D.R. et al., 1995, ApJ, 439, 933
\bibitem[\protect\citeauthoryear{Lyne}{2006}]{ly06}
Lyne, A.G., \& Graham-Smith, F. 2006(ed.), in Pulsar Astronomy (3rd ed., Cambridge Astrophysics Series; Cambridge Unive. Press), 264
\bibitem[\protect\citeauthoryear{Manchester}{2005}]{ma05}
Manchester, R.N., Hobbs, G.B., Teoh, A., Hobbs, M., Astron.
J., 129, 1993-2006 (2005) (astro-ph/0412641)
\bibitem[\protect\citeauthoryear{Manchester}{1999}]{ma99}
Manchester, R.N., 1999, in Pulsar Timing, General Relativity and the
Internal Structure of Neutron Stars, Edited by Z. Arzoumanian, F. Van
der Hooft, and E. P. J. van den Heuvel. Published
 by Koninklijke Nederlandse Akademie van Wetenschappen,
Amsterdam, The Netherlands, p. 53.
\bibitem[\protect\citeauthoryear{Narayan}{1990}]{na90}
Narayan, R. \&  Ostriker, J.P., 1990, ApJ, 352, 222
\bibitem[\protect\citeauthoryear{Paczynski}{1990}]{pa90}
Paczynski, B., 1990, ApJ, 348, 485
\bibitem[\protect\citeauthoryear{Ransom}{2011}]{ra11}
Ransom, S.M. et al., 2011, ApJL, 727, 16
\bibitem[\protect\citeauthoryear{Ray}{2010}]{ra10}
Ray P.S.,  2010, in ICREA Workshop on the High-Energy
  Emission from Pulsars and their Systems
\bibitem[\protect\citeauthoryear{Romani}{2010}]{ro10}
Romani R.W., Watter K.P., 2010, ApJ, 714, 810
\bibitem[\protect\citeauthoryear{Ruderman}{1991}]{ru91}
Ruderman M., 1991, ApJ, 366, 261	
\bibitem[\protect\citeauthoryear{Ruderman}{1975}]{ru75}
Ruderman M.A.,  Sutherland P.G., 1975, ApJ, 196, 51
\bibitem[\protect\citeauthoryear{Parkinson}{2010}]{pa10}
Saz Parkinson, P.M. et al. 2010, ApJ, 725..571
\bibitem[\protect\citeauthoryear{Shklovskii}{1970}]{sh70}
Shklovskii, I.S., 1970, Soviet Astron, 13, 562
\bibitem[\protect\citeauthoryear{Spitkovsky}{2006}]{sp06}
Spitkovsky, A.,  2006, ApJL 648, 51
\bibitem[\protect\citeauthoryear{Story}{2007}]{sto07}
Story, S.A.,  Gonthier, P.L., Harding, A.K., 2007, ApJ, 671, 713
\bibitem[\protect\citeauthoryear{Takata}{2010a}]{ta10a}
Takata J., Wang, Y.,  Cheng, K.S., 2010a, ApJ, 715, 1318
\bibitem[\protect\citeauthoryear{Takatab}{2010b}]{ta10b}
Takata J.,Cheng, K.S., Taam R.E., 2010b,  ApJL, accepted
\bibitem[\protect\citeauthoryear{Takata}{2011}]{ta11}
Takata J., Wang, Y.,  Cheng, K.S., 2011, ApJ, 726, 44
\bibitem[\protect\citeauthoryear{Venter}{2009}]{ve09}
Venter C.,  Harding A.K.,  Guillemot L., 2009, ApJ, 707, 800
\bibitem[\protect\citeauthoryear{Watters}{2011}]{wa11}
Watters, K.P., Romani, R.W., 2011, 727, 123
\bibitem[\protect\citeauthoryear{Zhang}{2003}]{za03}
Zhang L.,  Cheng K.S., 2003, A\&A 398, 639

\end{thebibliography}
\end{document}